\documentclass[twocolumn,prl,showpacs]{revtex4}

\usepackage{amssymb,epsfig}
\usepackage{algorithm}

\newcommand{\bra}[1]                             {{\langle#1|}}
\newcommand{\ket}[1]                             {{|#1\rangle}}

\newcommand{\Hil}                                {{\mathcal{H}}}
\newcommand{\nats}                               {{\mathbb N}}

\begin{document}

\title{Quantum Pattern Matching}

\author{P. Mateus}
\affiliation{Centro de L\'{o}gica e Computa\c{c}\~ao, Departamento
de Matem\'{a}tica, Instituto Superior T\'ecnico, P-1049-001
Lisbon, Portugal}
\author{Y. Omar}
\affiliation{Centro de F\'isica de Plasmas, Instituto Superior
T\'ecnico, P-1049-001 Lisbon, Portugal}

\date{31 August 2005}


\begin{abstract}
We propose a quantum algorithm for closest pattern matching which
allows us to search for as many distinct patterns as we wish in a
given string (database), requiring a query function per symbol of
the pattern alphabet. This represents a significant practical
advantage when compared to Grover's search algorithm as well as to
other quantum pattern matching methods \cite{ram:vin:03}, which
rely on building specific queries for particular patterns. Our
method makes arbitrary searches on long static databases much more
realistic and implementable. Our algorithm, inspired by Grover's,
returns the position of the closest substring to a given pattern
of size $M$ with non-negligible probability in $O(\sqrt{N})$
queries, where $N$ is the size of the string. Furthermore, we give
the full recipe to implement our algorithm (together with its
total circuit complexity), thus offering an oracle-based quantum
algorithm ready to be implemented.
\end{abstract}


\maketitle


Search in databases is nowadays a common and fundamental
application in computer science, one that we use daily to find a
word in a text, or a site in Google. Currently we are also living
the quantum information revolution, where the idea to encode
information in quantum systems offers us a radically new type of
information which allows for much securer communications and much
faster computations than what we were able to achieve so far using
(now-called) classical information \cite{ben:div:00, nie:chu:00}.
In particular, quantum cryptography has in recent years quickly
progressed from a ``beautiful idea" \cite{gis:02} to a
plug-and-play application that one can purchase. There have also
been a few quantum algorithms proposed (the most significant
probably being Shor's efficient factorization algorithm
\cite{sho:94,sho:96} of 1994, solving a problem that classically
is believed to be intractable) even though the construction of a
scalable quantum computer is still a challenge, presently being
tackled with a plethora of different technologies \cite{ARDA}.
Yet, should quantum computation become a reality, there is still
no implementable efficient quantum algorithm to search a given
database \footnote{Given the current models of quantum computer.},
despite Grover's celebrated quantum search algorithm proposed in
1996 \cite{gro:97}. Grover's work, which now constitutes a
paradigm for quantum search algorithms, offers a quadratic
speed-up in query complexity (i.e.\ calls of a query function)
when compared to the classical case. However, in the real
execution of these search algorithms, we must distinguish the
\emph{compile time} and the \emph{run time}. The compile time is
essentially the construction of the query function on which the
algorithm relies to identify the element being searched. But this
construction is, in general, not a negligible task. In particular,
for database search, we must go through all the database elements
to build the so-called oracle, so that we can then implement the
search. Note that this makes the quantum search irrelevant in
practical terms, since you need to know the solution to run it.
Moreover, given a query function built to find a particular
element, it can only be used again to find that very same element.
The search for a different item in the same database would require
building a new specific query function. All this represents a
serious obstacle to the application of current search algorithms
\footnote{Yet, Grover's algorithm can be extremely useful and
represent an effective speed-up for other search problems, as for
instance in checking if the elements of a list are solutions of a
given NP-complete problem, where the query function plays the role
of the verifier and can be easily implemented.}.

To address this problem, we propose a quantum algorithm for
pattern matching which allows us to search for as many distinct
patterns as we wish in a given unsorted string (database), and
moreover returns the position of the closest substring to a given
pattern with non-negligible probability in $O(\sqrt{N})$ queries,
where $N$ is the size of the string. This means that the time to
find the \emph{closest} match (a much harder problem than to find
the \emph{exact} match, as we shall see) does not depend on the
size of the pattern itself, a result with no classical equivalent.
Another crucial point is that our quantum algorithm is actually
useful and implementable to perform searches in (unsorted)
databases. For this, we introduce a query function per symbol of
the pattern alphabet, which will require a significant (though
clearly efficient) pre-processing, but will allow us to perform an
arbitrary amount of different searches in a static database. A
\emph{compile once, run many} approach yielding a new search
algorithm that not only settles the previously existing
implementation problems, but even offers the solution of a more
general problem, and with a very interesting speed-up. After
exposing in detail our algorithm and presenting the respective
analysis in the most significant limit (when the pattern is much
smaller than the text and not frequent), we give the explicit
recipe for the construction of the query functions and our
non-trivial initial state, including their circuit complexity
analysis. But let us start by briefly reviewing what is know
classically about the pattern matching problem.

In the classical setting, the best known algorithm for the closest
substring problem takes O$(M N)$ queries where $M$ is the size of
the pattern. This result follows from adapting the best known
algorithm for approximate pattern matching \cite{nav:01}, which
takes O($e N+M$) where $e$ is the number of allowed errors, and
take $e=(M-1)$, that is, the closest match could be a substring
that coincides just one letter with the pattern. One should not
compare the closest match to (exact) pattern match, where the
problem consists in determining if a certain word (pattern) is a
substring of a text. For exact pattern matching it is proven that
the best algorithm can achieve O($M+N$) \cite{nav:01}. However, in
practical cases where data can mutate over time, like DNA, or is
store in a faulty systems, the closest match problem is a much
more relevant, since sometimes, only approximates of the pattern
exist, but nevertheless need to be found.


Our algorithm is based on the modified Grover search algorithm proposed in
\cite{boy:bra:hoy:tap:98} for the case of multiple solutions. It uses the
techniques originally introduced by Grover \cite{gro:97}: a query operator
that marks the state encoding the database element being searched by
changing its phase; followed by an amplitude amplification of the marked
state. The state can be detected with non negligible probability by
iterating this process $\sqrt{N}$ times where $N$ is the size of the
database.

Let us now describe our closest pattern matching algorithm. Given a string
$w$ of size $N$ over an alphabet $\Sigma$, we want to know if a certain
pattern $p$ of size $M$ occurs in $w$, or at least obtain the closest
match to $p$ in $w$. In particular we want to find the position
$i\in\{1,\dots,N\}$ where a certain symbol of $p$ occurs in $w$. To this
end, we encode position $i$ in a unit vector $\ket{i}$ of a Hilbert space
$\Hil$ of dimension $N$ (where the set $B=\{\ket{1},\dots,\ket{N}\}$
constitutes an orthonormal basis of $\Hil$). Since we are considering
patterns of size $M$ the total search space will be $\Hil^{\otimes M}$.

The initial state of the total system reflects the fact that we
want the second symbol of $p$ to occur just after the first, and
the third to occur just after the second, and so on. For this
reason we consider the following initial entangled state, which
consists of a uniform superposition of all possible states
fulfilling this property:
\begin{equation}\label{initialstate}
\ket{\psi_0}=\sum_{k=1}^{N-M+1}\frac{1}{\sqrt{N-M+1}}\ket{k,k+1,\dots,k+M-1},
\end{equation}
thus restricting $\Hil^{\otimes M}$ to a subspace of dimension
$N-M+1$. $\ket{\psi_0}$ can easily be adapted to patterns with
gaps.

To perform the search, we now need to define a query operator $Q_\sigma$
 for each symbol $\sigma$ of the alphabet
$\Sigma$. We will thus have $|\Sigma|$ different query operators. Each
$Q_\sigma$  acts over $\Hil\otimes \Hil_2$ (where $\Hil_2$ is the Hilbert
space of dimension 2) as follows:
\begin{equation}
Q_\sigma({\ket{i}\otimes\ket{b}})={\ket{i}\otimes\ket{f_\sigma(i)\oplus
b}},
\end{equation}
where $\ket{i}$ encodes position $i$ and $\ket{b}$ is a auxiliary
qubit and $f_\sigma$ is a function such that:
\begin{equation}
f_\sigma(i)=
\left\{
\begin{array}{ll}%
1 & \textrm{if the $i$-th letter of $w$ is $\sigma$} \\
0 & \textrm{otherwise}
\end{array}\right..
\end{equation}
As in Grover's algorithm, we want to use the query to mark states
where there is a match for the individual symbol, in particular by
shifting the phase of the respective state, as given by the
following unitary transformation:
\begin{equation}
U_\sigma\ket{k}=(-1)^{f_\sigma(k)}\ket{k},
\end{equation}
where $\ket{k}\in B$.

\begin{figure}[ht]
\begin{center}
\epsfig{file=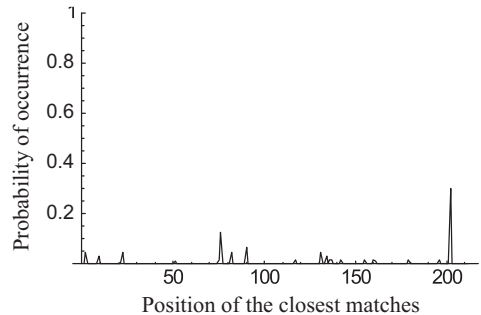, width=2.5in}
\end{center}
\caption{Simulation of our algorithm for a random string of size
$N=212$ and a particular pattern of size $M=10$ occurring towards
the end of the string.} \label{Fig simulation}
\end{figure}

However, in our quantum pattern matching algorithm a query
operator will be applied for a random symbol of the pattern to the
corresponding position. Hence, on average, a position with a
partial match, say of $M'$ out of $M$ matches of individual
symbols, will have the query operator applied $\frac{M'}{M}$
times. Note that the more matches we obtain, the more phase shift
will be shifted, and consequently the more the amplitude will be
amplified. Observe that for a given string there might be full and
partial matches, leading to larger and smaller amplitude
amplifications respectively (see Fig.\ \ref{Fig simulation} for an
example).
%

Note that, if N$>>$M, which is usually the case, sampling randomly over $M$
elements $\sqrt{N}$ times will lead to searching, with very highly probably,
over all elements of the pattern, that is, as $N$ grows, this probability
tends to 1 exponentially fast.

The amplitude amplification is obtained by applying the usual
Grover diffusion $D=D_N\otimes I^{\otimes M-1}$ to the total
state, where:
\begin{equation}
D_N=(2(\ket{\varphi}\bra{\varphi})-I),
\end{equation}
$I$ is the identity operator of dimension $N$ and $\ket\varphi \in\Hil$ is
given by the uniform superposition
$\ket\varphi=\sum_{i=1}^N\frac{1}{\sqrt{N}}\ket{i}$.

The algorithm is then constituted by iterating the phase shift induced by
the query followed by amplitude amplification. The final step is to
measure the state of a symbol of the pattern over the predefined basis
$B$, yielding the position of the closest match of the pattern in the
string. We show that it is enough to iterate $\sqrt{N}$ in order to
observe with non-negligible probability a match of the pattern.

In summary, the algorithm will be as follows:\\

\noindent
Input: $w\in \Sigma^*$ and $p\in \Sigma^*$\\ %
Output: $m\in\nats$ \\
Quantum variables: $\ket{\psi}\in \Hil(\{1,\dots,N\})^{\otimes M}$\\%
 Classical variables: $r,i,j\in \nats$
\begin{enumerate}
\item choose $r \in [0,\llcorner\sqrt{N-M+1}\lrcorner]$ uniformly,
\item set
$\ket{\psi}=\sum_{k=1}^{N-M+1}\frac{1}{\sqrt{N-M+1}}\ket{k,k+1,\dots,k+M-1}$;
\item for $i=1$ to $r$
\begin{enumerate}
\item choose $j\in[1,M]$ uniformly
\item set $\ket{\psi}=
            I^{\otimes j-1}\otimes Q^{w}_{p_j}\otimes
            I^{\otimes M-j} \ket{\psi}$;
\item set $\ket{\psi}=(D \otimes I^{\otimes M-1})\ket{\psi}$
\end{enumerate}
\item set $m$ to the result of the measurement of
the first component of $\ket{\psi}$ over the base
$\{\ket{1},\dots,\ket{N}\}$.
\end{enumerate}

The analysis of the algorithm follows closely that proposed in
\cite{boy:bra:hoy:tap:98}. The proof for the case where only exact
matches exist, and the symbols in the pattern occur only in these
matches, can be adapted straightforwardly, and will state that the
probability of finding a solution using the algorithm above is at
least $\frac{1}{4}$. When symbols occur elsewhere we are in the
context where the closest match, which we analyze next.

Assume that the alphabet is rich enough, and the symbols of the pattern do
not occur very often, if this is not the case one can combine letters in
pairs or triples. Moreover, if $N$ is very large and $N>>M$, then the
average amplitude is around $\frac{1}{\sqrt{N}}$. Note that, in this case,
each step of Grover amplification amplifies an amplitude $\alpha$ by
$\frac{2}{\sqrt{N}}$, since first it inverts the amplitude to $-\alpha$
and then applies the diffusion $D$ or inversion around average operator
that gives $\frac{2}{\sqrt{N}}+\alpha$. Now, if the pattern occurs in a
position $p$ then the random choice of $j$ at step 3 (a), will always lead
to an amplitude amplification of $p$. If there is a partial match, say
$M'$ symbols out of $M$, then, in average, the amplification will be done
$\frac{M'}{M}$ times.

We are now able to state the  amplitude amplification for a match of $M'$
out o $M$ is in average $\frac{M'}{M}$ times less than the amplification
for a perfect match, since $\frac{2}{\sqrt{N}}+\frac{2}{\sqrt{N}}$ is
added to the amplitude only $\frac{M'}{M}$ times.

Assuming an oracle for computing $Q^w_{p_j}$ for all $p_j$ in the
pattern, the query complexity of our pattern matching quantum
algorithm is $O(\sqrt{N})$, with no dependence on $M$, apart from
the cost of setting up of the initial state (\ref{initialstate}).
But to transform this interesting theoretical result into a useful
application, we now proceed to describe in detail how to build the
query functions and how to generate our non-trivial initial state,
thus giving the full recipe to implement our algorithm together
with its total circuit complexity.


The quantum circuit for the query operator is obtained from
implementing a permutation operator. As already noticed in
\cite{tof:80}, for any Boolean function of $n$ bits
$f:\{0,1\}^{n}\to \{0,1\}$ we are able to construct a bijection
$\tilde{f}$ on $n+1$ bits such that:
\begin{equation}
\begin{array}{ll}
\tilde{f}(0,x_1,\dots,x_n)=&
(f(x_1,\dots,x_n),x_1,\dots,x_n)\\
\tilde{f}(1,x_1,\dots,x_n)=& (1-f(x_1,\dots,x_n),x_1,\dots,x_n)
\end{array}.
\end{equation}
In the corresponding quantum case we have a Hilbert space
$\tilde{\Hil}$ of $n+1$ qubits and $\tilde{f}$ induces a unitary
transformation $U$ where:
\begin{equation}
U\ket{x_0,x_1,\dots,x_n}=\ket{\tilde{f}(x_0,x_1,\dots,x_n)}.
\end{equation}
Note that $U$ is the quantum implementation of the original Boolean
function $f$ we wish to calculate. The value of the function is stored in
the first qubit and the rest are ignored. Moreover, note that $U$ is
simply a permutation over the computational basis of $\tilde{\Hil}$, and
therefore can be obtained by composing $2^{n+1}-1$ transpositions (i.e.\
permutations of only two elements keeping the remaining unchanged), that
is, $U=U_1U_2\dots U_k$ with $k<2^{n+1}$ where $U_i$ acts only in two
elements of the basis.

Finally using Gray codes, we are able to implement each $U_i$
using at most $O(n^2)$ C-NOT and Pauli-$X$ gates. In detail,
recall that given two distinct binary words $\ell$ and $\ell'$ of
the same size $s$, a  Gray code from $\ell$ to $\ell'$ is a
sequence of binary words $r_0,\dots,r_k$ such that $r_0=\ell$,
$r_k=\ell'$ and $r_{j-1}$ differs only in one bit from $r_{j}$ for
any $j\in\{1,\dots,k\}$. Note that $k$ is less than or equal to
$s$, the size of the binary words. Given a Gray code we are able
to build the circuit to shift $\ell$ to $\ell'$, by applying in
sequence a controlled swap operation to the bit distinguishing
$r_{j-1}$ from $r_j$ for all $j\in \{1,\dots,k\}$. Albeit the
obtained permutation maps $\ket{\ell}$ to $\ket{\ell'}$, it is not
true in general that $\ket{\ell'}$ is mapped to $\ell$. Indeed,
$\ket{\ell'}$ is mapped to $\ket{r_{k-1}}$, and $\ket{r_j}$ is
mapped to $\ket{r_{j-1}}$ for any $j\in\{1,\dots,k\}$. In order to
obtain a transposition, we need to map $\ket{r_{k-1}}$ to
$\ket{\ell}=\ket{r_0}$ and $\ket{r_j}$ to $\ket{r_{j+1}}$ for all
$j\in{0,\dots,k-2}$. This can be achieved again by considering the
Gray code from $r_{k-1}$ to $r_0=\ell$, which is precisely
$r_{k_1},r_{k-2},\dots,r_0$. Again, by applying in sequence
controlled swap operation to the bit distinguishing $r_{j}$ from
$r_{j-1}$ for all $j\in \{k-1,\dots,1\}$ we attain the desired
transposition. Observe that for the particular case of
implementing the transposition $U_i$ on $\tilde{\Hil}$, the size
$s$ of the words is $n+1$.

In summary, given a transposition $U_i$ that transposes $\ket{\ell}$ with
$\ket{\ell'}$ and let $r_0,\dots r_k$ be the Gray code from $\ell$ to
$\ell'$, the algorithm implementing $U_i$ can be obtained as follows:\\

\noindent
Input: $\ket{\psi_0}\in \tilde\Hil$\\ %
Output: $U_i\ket{\psi_0}$\\ %
Classical variable: $i\in \nats$\\ %
Quantum variable: $\ket{\psi}\in \tilde\Hil$ %
\begin{enumerate}
\item swap $\ket{\psi}$ with $\ket{\psi_0}$
\item for $i=1$ to $k$
\begin{enumerate}
     \item set $\ket\psi=\texttt{C-NOT}(r_{i-1},r_i)\ket{\psi}$;
\end{enumerate}
\item for $i=k-1$ to $1$
\begin{enumerate}
     \item set $\ket\psi=\texttt{C-NOT}(r_{i-1},r_{i})\ket{\psi}$;
\end{enumerate}
\end{enumerate}
where the non-trivial gate $\texttt{C-NOT}(r_{i-1},r_{i})$ is the
transposition such that:
\begin{equation}
\begin{array}{ll}
\texttt{C-NOT}(r_{i},r_{i-1})\ket{r_i}=\ket{r_{i-1}}\\
\texttt{C-NOT}(r_{i},r_{i-1})\ket{r_{i-1}}=\ket{r_i}\\
\texttt{C-NOT}(r_{i},r_{i-1})\ket{w}=\ket{w}\textrm{ for all
}w\neq \ket{r_{i-1}},\ket{r_i}.
\end{array}
\end{equation}
In \cite{nie:chu:00} one can find a canonical construction of such
controlled gates requiring $O(n)$.

We conclude that any Boolean function of $n$ bits can be
implemented using $O(n^2 2^n)$ C-NOT gates and $O(2^n)$ Pauli-$X$
gates. This means that a query quantum circuit for inspecting a
list of $N$ elements can be built using $O( N
\log^2N)=\tilde{O}(N)$ gates.

The overall circuit complexity of our quantum pattern matching
algorithm --- excluding the initial setup --- is $O(
N^{3/2}\log^2(N)\log(M))$, versus $O(MN^2)$ for a classical
circuit (note that one needs $O(N)$ classical Boolean gates to
produce a circuit that  reads an arbitrary database of size $N$).

It remains to explain the cost of setting up the initial state
$\ket{\psi_0}$, given by equation (\ref{initialstate}), assuming
that initially all qubits are set to $\ket{0}$. Since we need $M$
variables ranging from $1$ to $N$, we will require $M\log(N)$
qubits to encode the quantum state of the program. We assume that
$N-M=2^s$ for some positive integer $s$ (if this is not the case,
we can augment the size of the string $N$ until this desideratum
is fulfilled and assume that no letter occurs in the augmented
part of the string).

\begin{figure}[ht]
\begin{center}
\epsfig{file=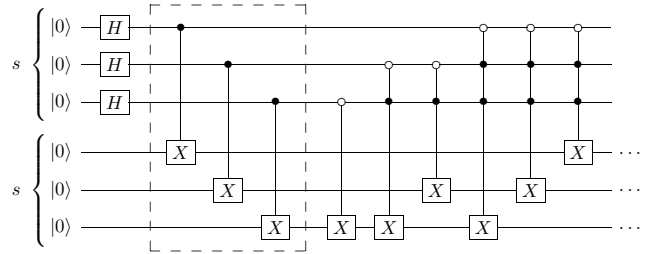, width=3.3in}
\end{center}
\caption{Core of the circuit to generate the initial state
$\ket{\psi_0}$ given by equation (\ref{initialstate}). The
first/second set of $s$ lines (in this example, $s=3$) represent
the qubits encoding the position of the first/second symbol of the
pattern. For the third symbol, we apply this same circuit
excluding the Hadamard operations to a new set of $s$ qubits,
controlled by the qubits of the second symbol, and so on. This
procedure must then be iterated another $M-3$ times, yielding an
overall complexity of $O(M\log^3(N-M))$.} \label{Fig circuit}
\end{figure}

We start by creating a uniform superposition of the $s$ qubits
encoding the position of the first symbol of the pattern $p$. This
is obtained by simply applying a Hadamard gate to each of these
qubits, as shown in Fig.\ \ref{Fig circuit} for $s=3$, and thus it
can be achieved in $O(s)$. The next step is to entangle these
qubits with the ones encoding the position of the second symbol of
$p$, and so on. We detail the process to do this for the second
symbol of $p$, and the final state is obtained by iterating this
process $M-2$ times.

First we create the state $\sum_{i=0}^{2^s-1}\ket{i,i}$, which can
be achieved by applying controlled Pauli-$X$ gates, as depicted in
the box of Fig.\ \ref{Fig circuit} for $s=3$. Finally, to obtain
the particular sequence encoding the order of the symbols of $p$,
we apply a sequence of $O(\log^2(N-M))$ multi-controlled Pauli-$X$
gates, as show in Fig.\ \ref{Fig circuit}. These multi-controlled
Pauli-$X$ gates can be implemented using  $O(\log(N-M))$ C-NOT and
Pauli-$X$ gates \cite{nie:chu:00}, and thus the overall circuit
complexity to construct this particular entanglement is
$O(\log^3(N-M))$.


The iteration is such that the $s$ qubits encoding the second
symbol of  $p$ are then used to control the qubits encoding the
third symbol of $p$, and so on. Hence, the overall circuit
complexity to construct the initial state given by equation
(\ref{initialstate}) is $O(M\log^3(N-M))$.

We conclude that our algorithm has an efficient compile time of
$O(N\log^2(N)\times|\Sigma|)$ and a total run time of
$O(M\log^3(N)+N^{3/2}\log^2(N)\log(M))$.



In summary, we have presented a quantum algorithm for closest
pattern matching that not only makes the quantum search in (long
unsorted) static databases realistic, but even interesting, as it
offers a faster solution than what is known classically for this
important problem. Based on a \emph{compile once, run many times}
approach, our algorithm allows for an arbitrary amount of
different searches on the same string, and offers a query
complexity of $O(\sqrt{N})$ in the most relevant limit where the
size $M$ of the pattern is much smaller than the size $N$ of the
database. Only the cost of setting up the initial state shows a
dependence on $M$. Furthermore, we gave the details of how to
obtain the full quantum circuit that implements our algorithm,
thus offering an oracle-based quantum algorithm ready to be
implemented.

\begin{acknowledgments}

The authors would like to thank A.\ Ambainis, L.\ Grover, E.\
Kashefi, J.I.\ Latorre, U. Vazirani and A.\ Sernadas for useful
discussions and remarks, and acknowledge the support from FCT and
EU FEDER through project POCI/MAT/55796/2004 QuantLog. YO would
also like to thank Funda\c{c}\~{a}o para a Ci\^{e}ncia e a
Tecnologia (Portugal) and the 3rd Community Support Framework of
the European Social Fund for financial support.

\end{acknowledgments}


\end{document}